\documentclass[a4paper, 11pt]{article}

\usepackage{graphicx}
\usepackage{ subfig}

\def\b{\begin{eqnarray}}
\def\e{\end{eqnarray}}
\def\g{\gamma}
\def\t{\tilde{\rho}}
\def\te{\tilde{\eta}}
\def\n{\noindent}
\begin{document}

\begin{center}

{\huge \textbf{Hamiltonian Dynamics of Cosmological \vskip.6cm Quintessence Models}}

\vspace {10mm}
\noindent
{\large \bf Rossen I. Ivanov and Emil M. Prodanov} \vskip.5cm
{\it School of Mathematical Sciences, Dublin Institute of Technology,
Ireland,} \vskip.1cm
{\it E-Mails: rossen.ivanov@dit.ie, emil.prodanov@dit.ie} \\
\vskip1cm
\end{center}

\vskip4cm
\begin{abstract}
\n
The time-evolution dynamics of two nonlinear cosmological real gas models has been reexamined in details with methods from the theory of Hamiltonian dynamical systems. These examples are FRWL cosmologies, one based on a gas, satisfying the van der Waals equation and another one based on the virial expansion gas equation. The cosmological variables used are the expansion rate, given by the Hubble parameter, and the energy density.
The analysis is aided by the existence of global first integral as well as several special (second) integrals in each case.
In addition, the global first integral can serve as a Hamiltonian for a canonical Hamiltonian formulation of the evolution equations.
The conserved quantities lead to the existence of stable periodic solutions (closed orbits) which are models of a cyclic Universe. The second integrals allow for explicit solutions as functions of time on some special trajectories and thus for a deeper understanding of the underlying physics. In particular, it is shown that any possible static equilibrium is reachable only for infinite time.
\end{abstract}

\newpage

\section{Introduction}

Quintessence is a dynamical, evolving, spatially-inhomogeneous component with negative pressure \cite{stein}, \cite{caldw}. It is characterised by an equation of state $p = \omega \rho$ linking the pressure $p$ to the energy density $\rho$ via the parameter $\omega$ which is a constant such that $-1 < \omega < - 1/3$ (the cosmological constant or vacuum energy is modelled by $\omega = -1$). Models for which $\omega < -1$, also characterised by negative pressure, are called phantom field models. For cosmic acceleration, it is required that $\omega$ must be smaller than $-1/3$ --- visible from the Friedmann equation $\ddot{a}/{a} = - (4 \pi G / 3)(\rho + 3p)$. This means that $\rho + 3p < 0$ --- a violation of the strong energy condition ($\rho + p \ge 0$ and $\rho + 3 p \ge 0$) \cite{car}. Any physical field with positive energy density (to account for  the necessary density to make the universe flat) and negative pressure (whether stemming from repulsive gravity or not), violating the strong energy condition, can play the role of dark energy \cite{leandros}. Commonly considered quintessential cosmological models are based on the introduction of a spatially-inhomogeneous slowly-evolving real scalar field rolling down a potential similar to the inflaton field in inflation theory. The pressure of the scalar field is negative if it rolls down so slowly that the kinetic energy density is smaller than the potential energy density.  Alternative quintessence models introduce real gas equations of state. Real gas equations of state have advantages over an ideal gas equation of state, since possible phase transitions between the thermodynamic states of cosmic fluids can be accounted for. In some epochs of the cosmological evolution, two phases could have existed together. \\
Phantom cosmological models violate all four energy conditions \cite{noj}. The phantom field is unstable from a quantum field theory perspective, but could be stable in classical cosmology \cite{noj}. A defining characteristic of these models is the so called Big Rip singularity --- the scale factor $a$ becoming infinite over a finite time (there are many proposed remedies for the avoidance of such singularity --- see \cite{noj} and the references therein). In his work \cite{caldwell2}, Caldwell introduces the concept of phantom fields by constructing a toy model of a "phantom" energy component which possesses an equation of state $\omega < - 1$ and arguing that it agrees, based on current data and understanding, with most classical tests of cosmology. In view of this, if future observations do not bar  $\omega < - 1$ models, the dominant component of the cosmic energy density may be very strange. Further, Carroll {\it et al.} argue \cite{car} that it is conceivable that a well-defined model could (perhaps temporarily) have $\omega < - 1$ , and indeed such models have been proposed. According to recent studies --- see \cite{leandros} and the references therein, phantom cosmologies are favoured over their quintessence counterparts. There is an ongoing discussion whether nature allows violation of all four energy conditions, even though it is very hard to make sweeping statements about what possible values $\omega$ may take and about a component of energy for which too little is known \cite{car}. It has been known for some time that such energy components can occur \cite {car}.  \\
Employing methods from the theory of Hamiltonian dynamical systems, presented in this paper is a  discussion of the solutions of two models, based on van der Waals and virial gas cosmologies, together with revealing of the Hamiltonian formulation of the nonlinear governing equations of these models and analysing the stable periodic solutions which are present (among others) in both. The Hamiltonian formulation is possible because of the existence of global first integrals in these non-linear models and such conserved quantities allow detailed analysis.  In particular, it leads to explicit solutions for special initial conditions, corresponding to special values of the conserved quantities. The cosmological variables used are the expansion rate, given by the Hubble parameter, and the energy density.  The former model is a quintessence one and is based on van der Waals real gas. It was originally proposed by Capozziello {\it et al.}  \cite{cap1}, \cite{cap2}, \cite{cap3} and has been studied further by many others. The latter model \cite{ivpro} replaces the van der Waals gas with a more general gas --- real virial gas --- and falls into the category of phantom field models.  The analysis of various mathematical aspects of these two models is quite interesting. \\
The paper is organised as follows: a brief cosmological set-up is followed by basic formulation of the real virial gas model and the van der Waals gas model. After some basic canonical Hamiltonian formulation tools are introduced, an illustration is made with a dynamic self-interaction model \cite{example} whose Hamiltonian structure has been revealed. The two main sections focus on the Hamiltonian formulations of the real virial gas and, separately, of the van der Waals gas, together with a thorough analysis of the trajectories in the phase plane, focusing in particular on closed curves which represent cyclic Universe scenarios.

\section{Real Virial Gas and van der Waals Gas}

The expansion of a homogeneous  Universe, modelled with a perfect fluid, is adiabatic (except in the early Universe when particle annihilation "pumps" heat and adiabaticity is temporarily lost). In general, a perfect fluid is characterised by equation of state of the type $p=p(\rho, T)$. However, in view of the adiabaticity, the fluid flow must be reversible (isentropic). This necessitates a {\it barotropic} equation of state $p=p(\rho)$ or motion of the fluid in a way that such relation effectively holds: only then the general case of two dynamical variables $\rho$ and $T$ reduces effectively to one \cite{ellis}. A barotropic fluid is an idealised situation and the relationship $p=p(\rho)$ is considered to be known in advance \cite{ogi}. Most of the important cosmological models, which include fluids and are dominant in different cosmological epochs, are barotropic. \\
If the gas is isothermal and ideal, then $p = c_s^2 \rho$ (where $c_s =$ const is the isothermal speed of sound). Alternatively, if the gas is isentropic and ideal, then $p = K \rho^c$ (where $K=$ const which ensures adiabaticity and $c$ is the ratio of the specific heats $C_p/C_v$) \cite{vtora}. \\
Following earlier work \cite{ivpro}, firstly, real gas is considered whose pressure $p$ is related to the particle number $N$, the temperature $T$, and the volume $V$ of the gas via the virial expansion
 \cite{mandl}:
 \b
 \label{vir}
 p = \frac{Nk_BT}{V} \Bigl[ 1 + \frac{N}{V} F(T) + \Bigl( \frac{N}{V} \Bigr)^2 G(T) + \cdots \Bigr].
 \e
The $F(T)$ term in this expansion corrects the ideal gas equation of state ($p = Nk_BT/V$) and is given by \cite{mandl}:
 \b
 F(T) = 2 \pi \int\limits_{0}^{\infty} \Bigl[1 - e^{-\frac{V(r)}{k_BT}} \Bigr] \, r^2 dr,
 \e
where $V(r)$ is the two-particle interaction potential. \\
Interactions involving three or more particles [the term $G(T)$ and beyond] will not be considered. \\
The two-particle interactions, which are slightly attractive at long distances and strongly repulsive at short range, are often viewed in  regularised form: ensemble of  identical ``hard spheres" of radius $a$, surrounded by square potential wells of width $ad$ ($d > 1$) and depth $- \epsilon$ (where $\epsilon > 0$). Namely, the two-particle interaction are given in regularised form by the potential
\b
 \label{potential}
 V(r) = \left\{
 \begin{array}{ll}
 \infty \, , & \mbox{$0 < r < a,$} \cr
 - \epsilon , &  \mbox{$a \le r \le ad, $}
 \cr  0 \, , & \mbox{$r > ad.$}
 \end{array}
 \right.
 \e
In Planck units ($k_B=1$), one finds the correction term $F(T)$ as:
\b
F(T) = 2 \pi \biggl[ \int\limits_{0}^{a} r^2 dr  \, + \, \int\limits_{a}^{ad} r^2 \Bigl( 1 - e^{\frac{\epsilon}{T}} \Bigr) dr \biggr] = \frac{2 \pi a^3}{3}
[ 1 +   (1 - e^{\frac{\epsilon}{T}}) (d^3 - 1)]  = - \alpha z(T),
\e
where  $z(T) = (e^{\frac{\epsilon}{T}} - 1) (d^3 - 1) - 1$ and $\alpha = (2/3) \pi a^3$. \\
In terms of the particle number density $n = N/V$, the equation of state of the real virial gas is:
\b
\label{eos}
p =  n T [1 - \alpha n z(T)].
\e
The relationship between the particle number density $n$ and the mass density $\rho$ can be established as follows:
$n = (N m) / (V m) = (M / V) (1/m) = \rho / m$, where $M$ is the mass of the system and $m$ is the relativistic mass of a representative particle: $m = m_0 + (1/2) m_0 u^2/c^2 + O(u^4/c^4).$ Here $m_0$ is the rest mass and $u$ is the speed of the particle. Thus, in units $c = 1$ one has: $\rho = n m = n m_0 + (1/2) n m_0 u^2 = \hbox{rest energy density} + \hbox{kinetic energy density}$. For monoatomic gas with three degrees of freedom --- all of which translational, the average kinetic energy is $(3/2) k_B T$ (with the units used, $k_B = 1$). Thus, the mass density can be written as $\rho = n m_0 + (3/2) n T$. If an ideal gas is considered, the pressure $p_0$ will be, according to the ideal gas law, $p_0 = n_0 T_0$ (quantities with index zero refer to an ideal gas). Thus, $p_0 = (2/3) (\rho_0 - n_0 m_0)$ or $\rho_0 = n_0 m_0 [1 + (3/2) (T_0/m_0)]$. For the electron, the rest mass $m_0$ is 511 keV or $10^9$ K approximately. That is, for quite high temperatures $T_0$, the term $T_0/m_0$ is quite small. That is, $p_0 = n_0 T_0$ is negligibly small and the energy density is mainly due to the rest energy density. When the random velocities, due to thermal agitation, are fully neglected, one is dealing with the called pressure-less (or dust) limit. \\
For the case of a real virial gas, the limit $T \ll m_0$ is not pressure-less: one should note that when $F(T) = 0$ (which happens at the so called Boyle temperature), the virial gas does indeed resemble an ideal gas mostly. However, below the Boyle temperature, the term $F(T)$ decreases without limit with the drop of the temperature towards 0. As will be shown, equilibrium points different from the origin, exist only for temperatures below the Boyle temperature, that is, interesting things occur below the Boyle temperature.
Thus taking a "dust" limit is merited. In such case:
\b
p = \rho \,\, \frac{\frac{T}{m_0}}{1 + \frac{3}{2} \frac{T}{m_0}} \Biggl[ 1 - \alpha \rho \,\, \frac{\frac{z(T)}{m_0}}{1 + \frac{3}{2} \frac{T}{m_0}}\Biggr] \approx \rho \tilde{T} [ 1 - \alpha \rho \tilde{z}(\tilde{T})],
\e
where $\tilde{T} = T/m_0$ is the new dimensionless temperature, and $\tilde{z}(\tilde{T})$ is obtained from $z(T)$ by replacing
$a$ with $\tilde{a} = (1/m_0)^{1/3} a, \,\,\epsilon$ with $\tilde{\epsilon} = \epsilon / m_0,$ and $T$ with $\tilde{T} = T/m_0$. The tildes will not be written from now on. \\
In the analysis, as in \cite{ivpro}, $m_0, \alpha, \epsilon,$ and $d$ will be the parameters of the model. Another parameter of the model will be temperature $T$. However, $T$ will be allowed to vary, and the different values of this parameter, would characterise different epochs and the evolution of the trajectories in the
phase-plane of the dynamical system. This means that a barotropic equation of state is considered, along the lines of that of an ideal gas, and does not mean that the temperature is forced to be constant.
The treatment of the situation is analogical to that in the standard case of  varying equation of state of an ideal gas: in principle, the ratio $\omega= p/\rho$ changes with time but it is assumed that any time derivatives of $\omega$ are negligible in comparison to those of $\rho$ --- a reasonable assumption given that the equation of state is derived micro-physically and is not linked to the expansion of the Universe.   \\
The Hamiltonian structure of this model will be revealed and analysis of the trajectories in the phase plane further studied. \\
In this work, the van der Waals quintessence scenario of Capozziello {\it et al.} \cite{cap1},
\cite{cap2}, \cite{cap3}, \cite{ivpro} will also be revisited and Hamiltonian formulation established. \\
The barotropic van der Waals equation of
state is \cite{mandl} $\Bigl[$see also \cite{cap1}, \cite{cap2}, \cite{cap3}, and \cite{ivpro}$\Bigr]$:
\b
\label{dw}
p = \frac{\g \rho}{1 - \beta \rho} - \alpha \rho^2 \, ,
\e
where $\alpha = 3 p_c/\rho_c^2$ and $\beta = 1/(3 \rho_c)$, with $\rho_c$ and $p_c$ being the density and pressure of the van der Waals gas at the critical point. Here $\g$ is the absolute temperature of the van der Waals gas. It will also be treated as a varying parameter of the model and it will be allowed to take negative values \cite{cap1}, \cite{cap2},
\cite{cap3}.

\section{Cosmological Setup}
This model in this paper follows that in \cite{ivpro} and describes the Universe classically as an infinite, flat, two-component mixture of baryonic dust  with energy density $\rho_b$ and pressure $p_b = 0$, and a real gas with equation of state derived from the real virial gas expansion (\ref{eos}) or the van der Waals model (\ref{dw}).  \\
Dynamical phase-plane analysis (with Hubble's parameter $H$ and density of the real gas
$\rho$ as dynamical variables) of a real virial gas model have shown \cite{ivpro} that there is initial data leading to a cyclic Universe solution that goes through an inflationary
phase in each cycle, together with open trajectories in the phase plane that may or may not pass through regions characterised by inflation. As the Universe cools down, the inflationary region on the phase plane decreases and eventually disappears in the limit $T \to 0$ \cite{ivpro}.  The cosmological model presented in \cite{ivpro} also does not exhibit an endless sequence of cycles of expansion and contraction.  The trajectory of the Universe on the
phase place is, in first approximation, an ellipse and the frequency of oscillations decreases, while the ratio of of its axes decreases as the
Universe is cooling with periodicity eventually
lost. \\
Cyclic solution also exists for the van der Waals model for ranges of the absolute temperature below zero \cite{ivpro}. \\
The set-up for the analysis of the two-fraction Universe --- for both types of gas --- is the
Friedmann--Robertson--Walker--Lema\^itre (FRWL) cosmology \cite{frwl} with flat spatial three-sections and metric:
\b
ds^2 = g_{\mu \nu} dx^\mu dx^\nu = dt^2 - a^2(t) [dr^2 + r^2 (d \theta^2 + \sin^2 \theta \, d \phi^2)],
\e
where $a(t)$ is the scale factor of the Universe.
\\
Geometrized units $c = 1 = G$ are used. \\
The matter energy-momentum tensor $T_{\mu \nu}$ is given by:
\b
\label{emt}
T_{\mu \nu} = (\tilde{\rho} + \tilde{p}) \, u_\mu \, u_\nu - \tilde{p} \, g_{\mu \nu} \, ,
\e
where $\tilde{\rho}$ and $ \tilde{p}$ are, respectively,  the cumulative density and pressure for both fractions and $u^\mu$ is the flow vector satisfying $g_{\mu \nu} u^\mu u^\nu = 1$. \\
It should be noted that real gases are legitimate perfect fluids, satisfying Euler equations, for as
long as dissipative forces are not included; namely, that there is no shear, stresses or heat conduction. Otherwise, a dissipative (or viscous) fluid (satisfying the Navier--Stokes equation) is characterised by a term additional to the ones already present in (\ref{emt}) --- the symmetric viscosity stress tensor $\sigma_{\mu \nu}$ (linearly perturbing the perfect fluid) \cite{y}:
\b
\sigma_{\mu \nu} = \lambda \pi_{\mu \nu} \nabla_\rho u^\rho + \nu (\nabla_\mu u_\nu + \nabla_\nu u_\mu),
\e
where the constants $\lambda$ and $\nu$ are the so-called bulk viscosity and shear viscosity, respectively, the projection tensor $\pi$ is given by $\pi_{\mu \nu} = g_{\mu \nu} + u_\mu u_\nu$. \\
Friedmann equations for the perfect fluid are \cite{friedmann}:
\b
\label{fr1}
\ddot{a} & = & - \frac{4 \pi}{3} (\tilde{\rho} + 3 \tilde{p})a, \\
\label{fr2}
\dot{a}^2 & = &  \frac{8 \pi}{3} \tilde{\rho} a^2
\e
or
\b
\label{h1}
H^2 & = & \frac{1}{3} (\rho_b + \rho), \\
\label{h2}
\dot{H} & = & - \frac{1}{2} (\rho_b + \rho + p)
\e
in terms of the Hubble parameter $H = \dot{a}/a$ (one of the two dynamical variables of the model). \\
The continuity equation for the real gas
\b
\label{cont}
\dot{\rho} + (\rho + p) \frac{3 \dot{a}}{a} = 0
\e
becomes
\b
\label{h3}
\dot{\rho} + 3H(\rho + p) = 0.
\e
The continuity equation for the pressure-less baryonic dust is:
\b
\label{h4}
\dot{\rho_b} + 3H \rho_b  = 0.
\e
Following \cite{ivpro}, differentiating (\ref{h1}) with respect to time and substituting into it $\dot{H}$ from (\ref{h2}),
$\dot{\rho}$ from (\ref{h3}) and $\dot{\rho_b}$ from (\ref{h4}), leads to an identity. Thus, equation (\ref{h1}) is just an integral of equations (\ref{h2}),
(\ref{h3}), and (\ref{h4}).  As it can be obtained from equations (\ref{h1}), (\ref{h2}) and (\ref{h3}), equation (\ref{h4}) will be dropped \cite{ivpro}.  \\
Expressing the baryonic energy density $\rho_b$ from equation (\ref{h1}) and substituting it into equation (\ref{h2}) gives the dynamical equation \cite{ivpro}:
\b
\label{hash}
\dot{H} = - \frac{3}{2} H^2 - \frac{1}{2} p.
\e
The other dynamical equation is (\ref{h3}) \cite{ivpro}:
\b
\label{rho}
\dot{\rho} = - 3H(\rho + p),
\e
with $\rho$ being the second dynamical variable. \\
Upon substitution of the equation of state (\ref{eos}), the dynamical system becomes \cite{ivpro}:
\b
\label{ddyn1a}
\dot{\rho} & = & - 3 H \rho \,  [1 + T - \alpha \rho T z(T)  ] \equiv f_1(\rho, H), \\
\label{ddyn1b}
\dot{H} & = & - \frac{3}{2} H^2 - \frac{1}{2} T \rho [ 1- \alpha \rho z(T) ]  \equiv f_2(\rho, H).
\e

\section{Canonical Hamiltonian Formulation}

Consider the following two-component autonomous system of ordinary differential equations,
\b
\dot{x} & = & f(x, y) \\
\dot{y} & = & g(x, y),
\e
where $f$ and $g$ are two $C^1$ functions for all $(x,y)\in \mbox{R \hskip-.5cm I\,\,\,}^2$. \\
It is also assumed that, in addition, a global first integral $I(x, y)=$ const,  exists, i.e.
\b
\label{idce}
\frac{\partial I}{\partial x} \dot{x} +
\frac{\partial I}{\partial y} \dot{y} \, = \,
\frac{\partial I}{\partial x} f(x, y)
+ \frac{\partial I}{\partial y} g(x, y) \, = \, 0.
\e
For simplicity, in the rest of this section it is further assumed that all introduced functions and inverse functions exist globally. In the following sections, any exceptions will be stated and dealt with separately. \\
In order to identify this integral as the Hamiltonian of the system, a change of variables is performed:
\b
\label{cha}
x & \rightarrow & p(x), \\
\label{chan}
y & \rightarrow & q(y),
\e
so that $\tilde{I}(p, q) = I \Bigl(x(p), y(q)\Bigr)$ satisfies:
\b
\dot{p} & = & - \frac{\partial \tilde{I}}{\partial q}, \\
\dot{q} & = & \frac{\partial \tilde{I}}{\partial p}.
\e
To determine the conditions under which this is possible, i.e. to see if such change of variables exists, consider:
\b
\dot{p} = \frac{dp}{dx}\frac{dx}{dt} = \frac{dp}{dx} f(x,y) & = & - \frac{\partial \tilde{I}}{\partial q} = - \frac{\partial I}{\partial y} \frac{dy}{dq} \\
\dot{q} = \frac{dq}{dy}\frac{dy}{dt} =\frac{dq}{dy} g(x,y) & = & \frac{\partial \tilde{I}}{\partial p} =  \frac{\partial I}{\partial x} \frac{dx}{dp}.
\e
Thus:
\b
\frac{dp}{dx} \frac{dq}{dy} & = & - \frac{1}{f(x, y)}\frac{\partial I}{\partial y}, \\
\frac{dp}{dx} \frac{dq}{dy} & = & \frac{1}{g(x, y)}\frac{\partial I}{\partial y}.
\e
The left-hand sides are the same and so are the right-hand sides --- in view of (\ref{idce}). \\
Therefore, any change of variables (\ref{cha}), (\ref{chan}) which satisfies the separability condition
\b
\label{sepa}
- \frac{1}{f(x, y)}\frac{\partial I}{\partial y} \equiv \frac{1}{g(x, y)}\frac{\partial I}{\partial x} = F(x) G(y),
\e
where $F(x) = dp/dx$ and $G(y) = dq/dy$, allows to identify $\tilde{I}(p, q)$ as the Hamiltonian of the system. \\
Of course, the identification of $p$ with the momentum and that of $q$ with the coordinate is only nominal, as any symplectic transformation, e.g. the change $p'=q, \, q'=-p$ also yields a canonical Hamiltonian system. \\
Additionally, for planar Hamiltonian systems, at equilibrium points, the eigenvalues of the linearised system are either purely real (i.e. $\pm \lambda$) or purely imaginary ($\pm i \lambda$). Thus, excluding the special case $\lambda_1 = 0 = \lambda_2$, which requires special treatment, the critical points are either saddles (the eigenvalues have opposite signs), or centres (the eigenvalues are purely imaginary). \\
To illustrate this, consider the planar Hamiltonian system
\b
\dot{p} & = & - \frac{\partial \mathcal{H}}{\partial q} \equiv u(p, q), \\
\dot{q} & = & \frac{\partial \mathcal{H}}{\partial p} \equiv v(p, q),
\e
Next, the dynamical system is linearised near an equilibrium point  $(p^\ast, q^\ast)$:
\b
\label{lin_dyna1}
\dot{p} & = & u(p, q) \,\, =  \,\, \Bigl(\frac{\partial u}{\partial p}\Bigr)^* (p - p^\ast) + \Bigl(\frac{\partial u}{\partial q}\Bigr)^*
(q - q^\ast) + \ldots, \\
\label{lin_dyna2}
\dot{q}  & = & v(p, q) \,\, =  \,\, \Bigl(\frac{\partial v}{\partial p}\Bigr)^*(p - p^\ast) + \Bigl(\frac{\partial v}{\partial q}\Bigr)^* (q - q^\ast) +
\ldots,
\e
where the stars on the derivatives indicate that they are taken at an equilibrium point $(p^\ast, q^\ast)$. In matrix form this can be written as:
\b
\label{ds1}
\frac{d}{dt} X(t) = L(p^\ast, q^\ast) \cdot X(t),
\e
where:
\b
\label{ds2}
X(t)= \left(
\begin{array}{c}
p(t) - p^\ast \cr
q(t) - q^\ast
\end{array}
\right)
\e
and the stability (Jacobian) matrix is:
\b
L(p^\ast, q^\ast) =
\left(
\begin{array}{cc}
\Bigl(\frac{\partial u}{\partial p}\Bigr)^\ast & \Bigl( \frac{\partial u}{\partial q} \Bigr)^\ast \cr \cr
\Bigl( \frac{\partial v}{\partial p} \Bigr)^\ast &  \Bigl( \frac{\partial v}{\partial q} \Bigr)^\ast
\end{array}
\right) =
\left(
\begin{array}{cc}
- \Bigl( \frac{\partial^2 \mathcal{H}}{\partial p \partial q} \Bigr)^\ast & - \Bigl( \frac{\partial^2 \mathcal{H}}{\partial^2 q} \Bigr)^\ast \cr \cr
\Bigl( \frac{\partial^2 \mathcal{H}}{\partial^2 p} \Bigr)^\ast & \Bigl( \frac{\partial^2 \mathcal{H}}{\partial p \partial q} \Bigr)^\ast
\end{array}
\right).
\e
As the trace of the stability
matrix is zero (i.e. $\lambda_1 + \lambda_2 = 0$), then $\lambda_1 \lambda_2 = -\lambda_1^2$ and the characteristic equation $\lambda^2 -($tr$L)\lambda + \det L = 0$ becomes simply $\lambda^2 - \lambda_1^2 = 0$. This yields eigenvalues given by $\pm \lambda_1$, if $\lambda_1$ is purely real (corresponding to a saddle point), or eigenvalues given by $\pm i \omega$, if $\lambda_1$ is purely imaginary (corresponding to a centre), or the special case $\lambda_1 = 0 = \lambda_2$. The eigenvalues cannot be  general complex numbers, as in such case the stability matrix cannot be traceless.\\
As an example of a Hamiltonian system, consider the dynamic self-interaction model \cite{example}:
\b
S[g_{ik}] = \int d^4 x \sqrt{-g} \Bigl[ \frac{R}{2 \kappa} + L_{(m)} + \phi(\Psi^2)\Bigr],
\e
where $R$ is the Ricci scalar, $L_{(m)}$ --- the matter Lagrangian, and $\phi(\Psi^2)$ is a function of $\Psi^2 = \Psi_{ik} \Psi^{ik}$ where $\Psi_{ik}$ are the components of the covariant derivative of the velocity four-vector $U_k$ of the matter: $\Psi_{ik} = \nabla_i U_k = U_i (U^m \nabla_m) U_k + \sigma_{ik} + \omega_{ik} + (1/3) \Delta_{ik} \Theta$, where
$\sigma_{ik} = (1/2) \Delta_i^m \Delta_k^n (\nabla_m U_n + \nabla_n U_m) - (1/3) \Delta_{ik} \Theta \,\, \omega_{ik} = (1/2) \Delta_i^m \Delta_k^n (\nabla_m U_n - \nabla_n U_m), \,\, \Theta = \nabla_m U^m,$ and the projector $\Delta_{ik}$ given by $g_{ik} - U_i U_k$. \\
For the metric considered in \cite{example}:
\b
ds^2 = dt^2 - a^2(t) (dx^2 + dy^2 + dz^2),
\e
the components of $\Psi_{ik}$ are given by:
\b
\Psi_{ik} = \nabla_i U_k = - \Gamma^0_{ik} = \frac{1}{2} \dot{g}_{ik} = \frac{1}{3} \Delta_{ik} \Theta = \frac{\dot{a}}{a} \Delta_{ik} = H(t) \Delta_{ik}.
\e
Thus $\Psi^2 = 3 H^2$. \\
The components of the energy-momentum tensor are given by \cite{example}:
\b
T^{(m)}_{ik} = W U_i U_k + \hat{P}_{ik},
\e
where $W$ is the energy density and the anisotropic pressure tensor $\hat{P}_{ik}$ can be decomposed as a sum of an isotropic part $\hat{p}$ and a non-equilibrium part $\Pi_{ik}$, that is: $\hat{P}_{ik} = \hat{p} \Delta_{ik} + \Pi_{ik}$. After eliminating the pressure with the barotropic relation $\hat{P} \equiv \hat{p} + \Pi = (\gamma - 1) W$, the dynamical equations of the model can be written as:
\b
\dot{H} & = & - \frac{1}{2} \, \frac{\kappa \gamma W}{1 + \frac{\kappa}{6} \phi''(H)} = f(H, W), \\
\dot{W} & = & - 3 \gamma H W = g(H, W).
\e
Thus, $dH/dW = \kappa [6H + \kappa H \phi''(H)]^{-1}$ or
$[6H + \kappa H \phi''(H)] dH - \kappa dW = 0$. \\
The conserved $I(H, W)$ is therefore given by:
\b
I(H, W) = 3 H^2 + \kappa [H \phi'(H) - \phi(H) - W] =
\mbox{const}.
\e
The separability condition (\ref{sepa}) is satisfied:
\b
\label{heree}
-\frac{1}{g(H,W)}\frac{\partial I}{\partial H}= \frac{1}{f(H,W)}\frac{\partial I}{\partial W}
 & = & \frac{1}{3 \gamma H W} [6 H + \kappa H \phi''(H)] \nonumber \\
& = & [2 + \frac{\kappa}{3} \phi''(H)] \, \frac{1}{\gamma W} = F(H) G(W).
\e
To identify the integral with the Hamiltonian, a change of variables $H \rightarrow q(H)$ and $W \rightarrow p(W)$, as suggested from (\ref{heree}), is performed so that:
\b
\frac{dp}{dW} & = & \frac{1}{\gamma W}, \\
\frac{dq}{dH} & = &
2 + \frac{\kappa}{3} \phi''(H).
\e
Therefore, in variables
\b
p & = & \frac{1}{\gamma} \ln W, \\
q & = & 2 H +\frac{\kappa}{3} \phi'(H),
\e
the model has a canonical Hamiltonian formulation. \\
The explicit map to the canonical coordinates $(p, q)$ shows that the system is globally Hamiltonian. The new coordinates $(p,q)$ provide explicitly the so-called Darboux chart for the configuration manifold, where the symplectic structure acquires the standard form $\omega = dq \wedge dp$,  see more details in \cite{books}. \\
The existence of such coordinates is a general fact: as dynamical systems in I$\hskip-2pt\mathrm{R}^2$ with a global first integral have one degree of freedom and as any differential 1-form in I$\hskip-2pt\mathrm{R}^2$ admits an integrating factor, then any such dynamical system will always be Hamiltonian \cite{books}. A very interesting example of a globally Hamiltonian system arising in cosmology (Einstein Static Universe in Massive Gravity) is studied in details in \cite{parisi}.

\section{Hamiltonian Formulation and Dynamics of a Cosmological Model with Virial Gas}

The dynamical equations (\ref{ddyn1a}) and (\ref{ddyn1b}) of the real virial gas system can be written as:
\b
\label{dyn1a}
\dot{H} & = & - \frac{3}{2} H^2 - \frac{1}{2} T \rho [ 1- \alpha \rho z(T) ] = -\frac{3}{2} H^2 + \frac{1}{2} [\rho - b(\rho)], \\
\label{dyn1b}
\dot{\rho} & = & - 3 H \rho \,  [1 + T - \alpha \rho T z(T)] = -3Hb(\rho),
\e
where $b(\rho) =
 \rho \,  [1 + T - \alpha \rho T z(T)]$. \\
A constant $I(H, \rho)$ is souught such that:
\b
0 = \frac{dI}{dt} = \frac{\partial I}{\partial H} \dot{H} + \frac{\partial I}{\partial \rho} \dot{\rho}
\e
or, using the dynamical equations (\ref{dyn1a}), (\ref{dyn1b}),
\b
\frac{\partial I}{\partial H}
\Bigl[-\frac{3}{2} H^2 + \frac{1}{2} [\rho - b(\rho)]
\Bigr]
+ \frac{\partial I}{\partial \rho} [-3 H b(\rho)] = 0.
\e
Now change variables $\rho \rightarrow \t(\rho)$ via:
\b
\label{changevar}
- b(\rho) \frac{\partial}{\partial \rho} = \t \frac{\partial}{\partial \t}.
\e
Then $I[H, \rho(\t)]$ satisfies:
\b
\frac{\partial I}{\partial H}
\biggl[-\frac{3}{2} H^2 + \frac{1}{2} \Bigl[\rho(\t) - b[\rho(\t)]\Bigr]
\biggr]
+ \frac{\partial I}{\partial \t} [3 H \t] = 0.
\e
Thus $I[H, \rho(\t)]$ can be found as a solution of the system:
\b
\label{gg1}
\frac{\partial I}{\partial \t} & = & -\frac{3}{2} H^2 + \frac{1}{2} \Bigl[\rho(\t) - b[\rho(\t)]\Bigr], \\
\label{gg2}
\frac{\partial I}{\partial H}
& = & - 3 H \t.
\e
(Indeed, the mixed derivatives are equal: $\partial_{\t} \partial_H I = \partial_H \partial_{\t} I = - 3H$.) \\
The integral $I[H, \rho(\t)]$ can be identified with the canonical Hamiltonian since the Hamilton equations are satisfied:
\b
\label{ham}
\dot{H} & = & \frac{\partial I}{\partial \t}, \\
\dot{\t} & = & - \frac{\partial I}{\partial H}.
\e
Integrating (\ref{changevar}) gives:
\b
\label{rt}
\t = e^{-\int \frac{d \rho}{b(\rho)}} = \Bigl[ \frac{1+T}{\rho}  - \alpha T z(T)\Bigr]^{\frac{1}{1+T}}.
\e
Thus:
\b
\rho = \frac{1+T}{\t^{1+T} + \alpha T z(T)}.
\e
One should note that $\t \to \infty$ as $\rho \to 0$. \\
Integrating (\ref{gg1}) with respect to $\t$ gives:
\b
\tilde{I}[H, \tilde{\rho}] \equiv
I[H, \rho(\t)] =
- \frac{3}{2} H^2 \t + \frac{1+T}{2}
\frac{\t}{\t^{1+T} + \alpha T Z(T)}.
\e
In terms of the original variables, the Hamiltonian is:
\b
\label{hamor}
I(H, \rho) = - \frac{1}{2} \Bigr[ \frac{1+T}{\rho} - \alpha T z(T) \Bigr]^{\frac{1}{1+T}} (3 H^2 - \rho).
\e
The origin $(\rho^\ast_1 = 0, \,\, H^\ast_1 = 0)$ of the phase-plane is an equilibrium point since $\dot{\rho} = 0 = \dot{H}$ there \cite{ivpro}.
The change of variables into canonical Hamiltonian variables (\ref{rt}) moves the equilibrium $(0,0)$ to $(\infty,0)$, and thus, strictly speaking, this point is not "visible" in the new coordinates. \\
Other equilibrium points can exist only for negative pressure $p$ \cite{ivpro}: in view of (\ref{hash}), at an equilibrium point, $\dot{H} = 0$, thus, if $H^\ast \ne 0$ at that point, then $p < 0$ \cite{ivpro}. The pressure $p$ is negative when:
\b
\label{neg}
1 - \alpha \rho z(T)   < 0
\e
(the parameter $\alpha$ and the density $\rho$ are both positive). \\
Thus equilibrium points, for which $H^\ast \ne 0$, can occur only for values of the density $\rho$  greater  \cite{ivpro} than
\b
\label{rmin}
\rho_{\mbox{\tiny min}} = \frac{1}{\alpha z(T)} = \frac{1}{ \alpha [(e^{\frac{\epsilon}{T}} - 1) (d^3 - 1) - 1 ]  }.
\e
For $\rho_{\mbox{\tiny min}}$ to be positive, $z(T) = (e^{\frac{\epsilon}{T}} - 1) (d^3 - 1) - 1$ must be positive. That is, there is an upper limit on the temperature below which equilibrium points, different from the origin, exist and this upper limit is the Boyle temperature:
\b
T_{\mbox{\tiny max}} = T_B =  \frac{\epsilon}{\ln d^3 - \ln (d^3 -1)}.
\e
The other equilibrium points are \cite{ivpro}  point $Q$ with coordinates
$(\rho^\ast_2 = \rho_{\mbox{\tiny min }}, \,\, H^\ast_1 = 0)$, point $R$ with coordinates $ (\rho^\ast_3 \, , H^\ast_{2})$ and point $S$ with coordinates $(\rho^\ast_3 \, , H^\ast_{3})$ (see Figure 1) where:
\b
\label{r2}
\rho^\ast_3 \,\, = \,\, \frac{1+T}{\alpha T z(T)}
\,\, = \,\, \rho_{\mbox{\tiny min}} (1 + \frac{1}{T}).
\e
and
\b
\label{h23}
H^\ast_{2,3} \,\, = \,\, \pm \,\, \sqrt {\frac{\rho^\ast_3}{3}} \,\,
=  \,\, \pm \,\, \sqrt {\frac{1}{3} \rho_{\mbox{\tiny min}} (1 + \frac{1}{T})}.
\e
The stability matrix $L(\rho, H)$ is \cite{ivpro}:
\b
\label{L}
L(\rho, H) = \left(
\begin{array}{cc}
- 3 H(1+T) + 6H \alpha T \rho z(T) & - 3 \rho [1 + T - \alpha T \rho z(T) ]
\cr
- \frac{T}{2} + \alpha T \rho z(T) & -3H
\end{array}
\right).
\e
At the origin $(\rho^\ast_1 = 0 \, , H^\ast_{1} = 0), \,\,
\lambda = 0$ is a double eigenvalue and to analyse the situation, the conserved quantity $I$ will be used. \\
At the equilibrium point $R$ with coordinates $(\rho^\ast_3 \, , H^\ast_{2})$ [determined in (\ref{r2})--(\ref{h23}) above], the eigenvalues of
$L(\rho^\ast_3 \, , H^\ast_{2})$ are \cite{ivpro}:
\b
\lambda_1 & = & \sqrt{3 \rho^\ast_3} \,\,\, (1 + T) =  3 H_2^\ast \, (1+T) > 0, \\
\lambda_2 & = & - \sqrt{3 \rho^\ast_3} = -3 H_2^\ast < 0.
\e
As the eigenvalues have opposite signs, this equilibrium point is a saddle point \cite{ivpro}.   \\
Similarly, equilibrium point $S$ with coordinates $(\rho^\ast_3 , H^\ast_3)$ is another saddle point (both eigenvalues flip their signs): $\lambda_1 =  3 H_3^\ast \, (1+T) < 0$ and $\lambda_2 = -3 H_3^\ast > 0$ \cite{ivpro}.  \\
Equilibrium point $Q$ with coordinates $(\rho^\ast_2 = \rho_{\mbox{\tiny min}} \, , H^\ast_{1} = 0)$ is a centre (the eigenvalues there,
$\lambda^2 = - (3T)/[2 \alpha z(T)]$, are purely imaginary) \cite{ivpro}.
\begin{center}
\includegraphics[width=6cm]{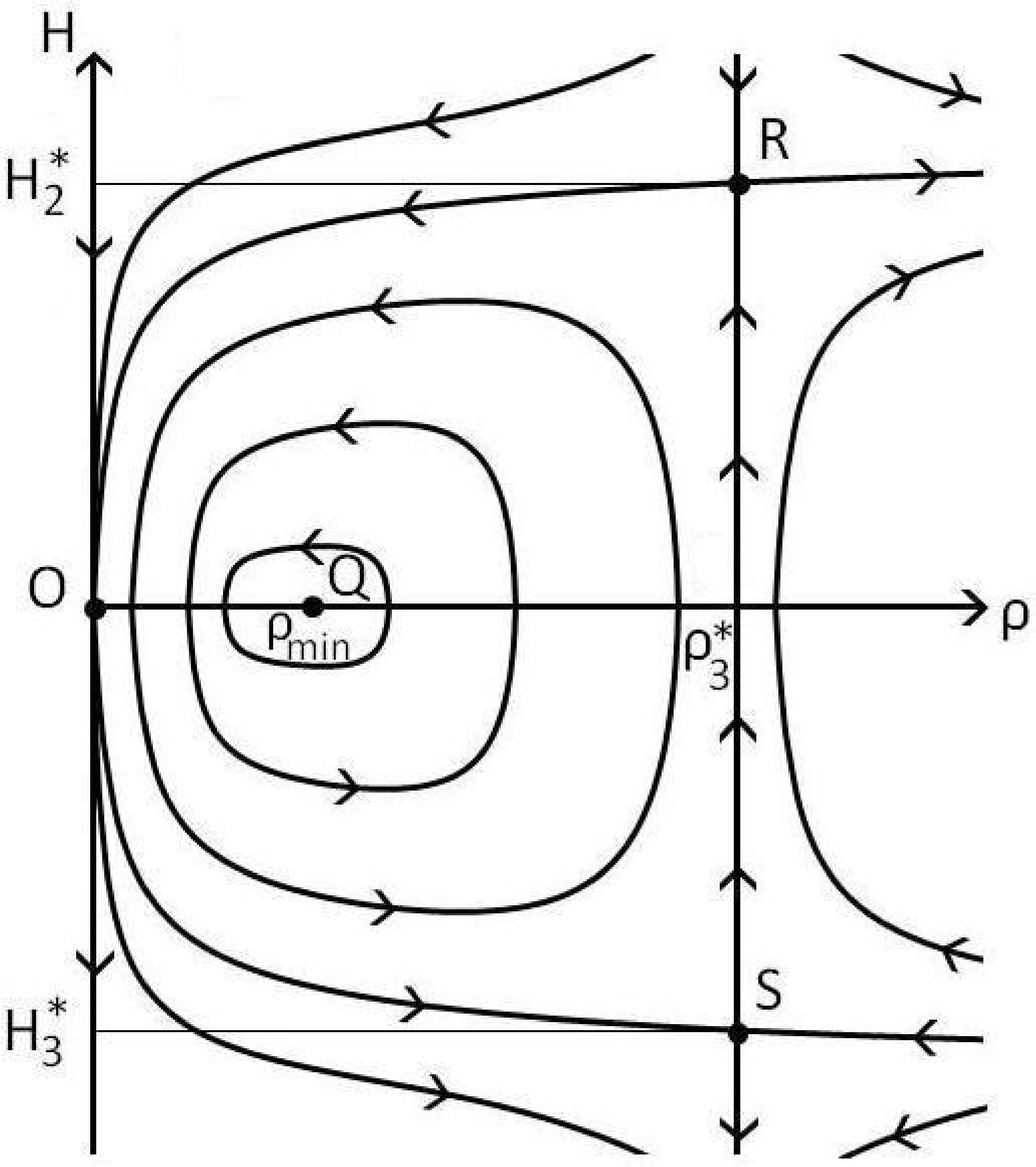}
\vskip.3cm
\noindent
\parbox{100mm}{\footnotesize {\bf Figure 1}: Equilibrium points for real virial gas model for temperatures below the Boyle temperature.}
\vskip.4cm
\end{center}
In order to study the phase-plane trajectories near the centre $Q$ (see Figure 1), the Hamiltonian (\ref{hamor}) is expanded in power series near point $(\rho^*_2, 0)$. That is, at point with coordinates
$(\rho=\rho^\ast_2+r = \rho_{\mbox{\tiny min}} + r, \quad  H=H^\ast_{1}+h = 0 + h)$, where $r$ and $h$ are infinitesimal, i.e. near the centre $Q$, the canonical Hamiltonian, up to and including the quadratic terms, is:
\b
\label{hamex}
I(h,\rho_{\mbox{\tiny min}} + r ) =
\frac{1}{2} \rho_{\mbox{\tiny min}}^{\frac{T}{T+1}} +
 \Bigl( \frac{1}{\rho_{\mbox{\tiny min}}}\Bigr)^{\frac{1}{1+T}}
\Bigl( -\frac{3}{2} h^2 + \frac{1}{2} \rho_{\mbox{\tiny min}} -\frac{T}{4} \frac{1}{\rho_{\mbox{\tiny min}}} r^2 \Bigr) = \mbox{const.}
\e
Thus
\b
\label{elipsa}
\frac{6 \rho_{\mbox{\tiny min}}}{T} h^2 + r^2 = \mbox{const.}
\e
The trajectory is an ellipse:
\b
r & = & C \cos \omega t, \\
h & = & C \sigma \sin \omega t,
\e
where $C$ is a constant depending on the initial conditions, $\omega = [3T/(2 \alpha z(T))]^{1/2}$ is the angular frequency of the oscillations and $\sigma = [\alpha T z(T)/6]^{1/2}$ is the ratio of the axes of the ellipse. \\
As noted in \cite{ivpro}, for the cooling Universe, with the drop of the temperature ($T \to 0$), the angular frequency $\omega$ decreases to zero and, in result, the period of oscillations increases indefinitely and thus periodicity is lost, i.e. only the hot Universe is cyclic. At the same time, the ratio $\sigma$ increases to infinity. \\
This model is also characterised by inflation: the region in the upper half of the phase plane ($H > 0$) for which:
\b
H^2 < \alpha T z(T) \Bigl(\rho - \frac{\rho_{\mbox{\tiny min}}}{2}\Bigr)^2 - \frac{1}{4} \frac{T}{\alpha z(T)}.
\e
is inflationary \cite{ivpro}.  This region is bounded by a hyperbola with asymptotes \cite{ivpro}
\b
H = \pm \sqrt{\alpha T z(T)} \Bigl(\rho - \frac{\rho_{\mbox{\tiny min}}}{2}\Bigr)
\e
and cutting the $\rho$-axis exactly at $\rho = \rho_{\mbox{\tiny min}} = [\alpha z(T)]^{-1}$. \\
With the drop of the temperature ($T \to 0$), the angle between the asymptotes
decreases to zero and the inflationary regime is eventually switched off \cite{ivpro}.  \\
The second integrals of this Hamiltonian system are also of interest. A second integral, $K(\vec{x})$, is an invariant, but only on a restricted subset, given by its zero level set. It is defined by $\dot{K}(\vec{x}) = \mu(\vec{x}) K(\vec{x})$ \cite{gor}. Second integrals can neither predict the existence of first integrals, nor are able to give a global picture of the phase portrait \cite{gor}. They reduce to first integrals when $\mu = 0$ and to time-dependent first integrals when $\mu = $const \cite{gor}. Second integrals were studied by Darboux, Poincar\'e, Painlev\'e, and others \cite{gor}. \\
From (\ref{hamor}), it is clear that the parabola $\rho = 3 H^2$ is a special curve and, despite the fact that the first integral $I(H, \rho)$ is not defined at the origin, it is zero everywhere on the parabola. Also, $K_1 = \rho - 3 H^2$ is a second integral. \\
The dynamical equations (\ref{dyn1a}) and (\ref{dyn1b}) are automatically satisfied on the parabola $\rho = 3 H^2$. \\
The eigenvectors of the stability matrix (\ref{L}) at points $(\rho^\ast_3 \, , H^\ast_{2,3})$, corresponding to eigenvalue $\lambda_1 = - 3 H_{2,3}^\ast$ is $\vec{u}_1 = (0, 1)$ and the eigenvectors at points $(\rho^\ast_3 \, , H^\ast_{2})$, corresponding to eigenvalue $\lambda_2 = 3 H^\ast_{2,3}$ are $\vec{u}_2 = (6H_{2,3}^\ast, 1)$. The latter are tangent to the parabola at the saddle points [the slope of the parabola, $dH/d\rho$ at the saddles is $1/(6H_{2,3}^\ast)$]. \\
One should observe that $ROS$ is a heteroclinic orbit (it goes through the equilibria $R$, 0, $S$). All orbits inside the heteroclinic orbit are with equation $I(H, \rho) =$ const and are closed (cyclic Universe). All other trajectories are unbounded.\\
Integrating the dynamical equation (\ref{dyn1a}) on the parabola $\rho = 3H^2$ yields:
\b
\frac{1}{H_2^*} \,\,
\ln \Biggl| \frac{1-\frac{H}{H_3^*}}{1-\frac{H}{H_2^*}} \, \frac{1-\frac{H_0}{H_2^*}}{1-\frac{H_0}{H_3^*}} \Biggr|
- 2\Bigl( \frac{1}{H} - \frac{1}{H_0}\Bigr)
 = - \,\, 3(1+T) (t-t_0),
\e
where $H_0 = H(t_0)$. \\
It is clear from here that when $H \to H^\ast=0$, the term $1/H$ blows up. Therefore, $H^\ast = \pm 0$ is reachable in infinite (reversed) time ($t \to \pm \infty$) while on the parabola $\rho = 3H^2$.
When $H = H_3^\ast$, that is, moving towards the saddle $S$, infinite time is needed ($t \to \infty$) to reach $S$.  Similarly, return to the saddle $R$ (i.e. $H = H_2^\ast$) along the parabola will take infinite reversed time ($t \to - \infty$). \\
The straight line joining the saddles $S$ and $R$ is with equation $\rho = \rho_3^\ast =$ const.  Another second integral, conserved only on this line, is $K_2 = \rho - \rho_3^*$. \\
Integrating the dynamical equation (\ref{dyn1a}) on this straight line yields:
\b
\ln \Biggl| \frac{1-\frac{H}{H_3^*}}{1-\frac{H}{H_2^*}} \, \frac{1-\frac{H_0}{H_2^*}}{1-\frac{H_0}{H_3^*}} \Biggr|
 =  3 H_2^* (t-t_0).
\e
Again, the saddles are reachable in infinite (reversed) time ($t \to \pm \infty$) when moving on the vertical line $\rho = \rho_3^\ast \, = $ const.  \\
Finally, consider the ordinate $\rho = 0$. The quantity $K_3 = \rho$ is also another second integral, conserved only on $\rho=0$. \\
Integrating equation (\ref{dyn1a}) results in:
\b
\frac{1}{H} = \frac{1}{H_0} + \frac{3}{2} (t -t_0).
\e
Taking $H \to \pm 0$ shows that the origin is also reachable in infinite (reversed) time ($t \to \pm \infty$) along the straight line $\rho = 0$.\\
In terms of the canonical variables $H$ and $\t$ in (\ref{rt}), the centre and the two saddles are preserved, i.e. they are at $\Bigl(\t^*=[\alpha z(T)]^{\frac{1}{1+T}},$ $H^*=0\Bigr)$ (the centre) and at $\Bigl(\t^*=0, $  $H^*=\pm\sqrt{\rho_3^*/3}\Bigr)$ (the two saddles).

\section{Hamiltonian Formulation and Dynamics of a Cosmological Model with van der Waals Quintessence}
\n
In terms of the dimensionless energy density $\eta$, defined via $\eta = \rho/\rho_c > 0$ [where the critical density  $\rho_c$ is  $3 H_0^2/(8 \pi G)$)],  the van der Waals equation of state (\ref{dw}) is  \cite{cap1}, \cite{cap2}, \cite{cap3}, \cite{ivpro}:
\b
\label{eeta}
p = \frac{3 \g \rho}{3 - \eta} - \frac{9}{8} \g \eta \rho \, .
\e
Here $\gamma$ is the absolute temperature, which will be allowed to take negative values. The temperature will be treated as a parameter. \\
Substituting the van der Waals equation of state (\ref{eeta}) into the dynamical equations (\ref{hash}) and (\ref{rho}) gives \cite{ivpro}:
\b
\label{dyn2}
\dot{\eta} & = & - 3 H \eta \,  (1 + \frac{3 \g}{3 - \eta} - \frac{9}{8} \g \eta), \\
\label{dyn1}
\dot{H} & = & - \frac{3}{2} H^2 - \frac{8 \xi \g \eta }{3 - \eta} + 3 \xi \g \eta^2,
\e
where $\xi = (3/16)  \rho_c = $ const $>0$ is another parameter of the model. \\
The equilibrium points are determined by requesting $\dot{H} = 0$ and $\dot{\eta} = 0$ and the origin $(\eta=0, H = 0)$ is immediately identifiable as an equilibrium point. Next, requesting the term in the brackets in  (\ref{dyn2}) to be zero, results in the following quadratic equation \cite{ivpro}:
\b
\label{roots}
\frac{9}{8} \g \eta^2 - (\frac{27}{8} \g  + 1) \eta + 3(1 + \g) = 0 \, .
\e
For real roots
\b
\label{eta_roots}
\eta_{1, 2}^\ast(\g) = \frac{\frac{27}{8} \g  + 1 \pm \sqrt{- \frac{135}{64} \g^2 - \frac{27}{4} \g + 1 }}
{\frac{9}{4}\g}
\e
to exist, the discriminant
\b
\label{discr}
D = - \frac{135}{64} \g^2 - \frac{27}{4} \g + 1
\e
must be positive. That is, the range of values of $\g$, allowing two real solutions $\eta_{1,2}^*$ to exist at which $\dot{\eta}=0$ and which are away from the origin, is:
\b
\label{gamma}
-\frac{8}{5} - \frac{32}{45} \sqrt{6} < \g < -\frac{8}{5} + \frac{32}{45} \sqrt{6}
\e
or $-3.3419 < \g < 0.1419$.\\
As $\eta_{1,2}^*$ are energy densities, they cannot be negative. In the sub-interval $-1 < \g < 0$, there is only one physically meaningful root: $\eta_1^*$. The other root, $\eta_2^*$, is negative. \\
Looking at the other dynamical equation, (\ref{dyn1}), there are four values for $H$ (two in the case when $-1 < \g < 0$) which satisfy $\dot{H}=0$ \cite{ivpro}:
\b
\label{hash_roots}
H^\ast_{1,2}(\g) & = & \sqrt{\frac{2 \xi \eta_{1,2}^\ast}{3}}\sqrt{\frac{\g (3 \eta_{1,2}^{\ast^2} - 9 \eta_{1,2}^\ast + 8)}{\eta_{1,2}^\ast - 3}}, \\
H^\ast_{3,4}(\g) & = & - H^\ast_{1,2}(\g).
\e
For these to be real, the following must hold:
\b
\g (\eta_{1, 2}^\ast - 3) (3 \eta_{1, 2}^{\ast^2} - 9 \eta_{1, 2}^\ast + 8) > 0 \, .
\e
As $3 \eta_{1, 2}^{\ast^2} - 9 \eta_{1, 2}^\ast + 8 = (\eta_{1, 2}^\ast - 3/2)^2 + 5/12 > 0,$  taking into consideration (\ref{gamma}), there are two
regimes: $-8/5 - (32/45) \sqrt{6} < \g < 0$ (in which case it follows that $\eta_{1,2}^\ast < 3$)
and $0 < \g < -8/5 + (32/45) \sqrt{6}$ (which leads to $\eta_{1,2}^\ast > 3$). \\
In particular, as $\g$ varies over the entire interval: $ -8/5 - (32/45) \sqrt{6} < \g < -8/5 + (32/45) \sqrt{6}$
(or $-3.3419 < \g < 0.1419$), the root $\eta_1^*(\g) = [(4 \g)/9]
\Bigl[(27/8) \g + 1 + [-(135/64) \g^2 - (27/4) \g + 1 ]^{1/2}\Bigr]$ is finite and varies between $3 - (2/3) \sqrt{6}$ and $3 + (2/3) \sqrt{6}$, that is, between 1.3670 and 4.6330. \\
On the other hand, when $-3.3419 < \g < -1$, the root $\eta_2^*(\g) = [(4 \g)/9]
\Bigl[(27/8) \g + 1 - [-(135/64) \g^2 - (27/4) \g + 1 ]^{1/2}\Bigr]$, drops from 1.3670 to 0. When $-1 < \g < 0$, $\eta_2^*$ becomes negative and tends to $-\infty$ as $\g \to 0^-$. In this situation, this root is unphysical and has to be discarded. When $0 < \g < 0.1419$, the root $\eta_2^*$ drops from $+ \infty$ to 4.6330. \\
The eigenvalues of the matrix $L(\eta, H)$ of the linearised dynamical system are \cite{ivpro}:
\b
\label{l1}
\lambda_1 & = & - 9 \g H^\ast_{1,2,3,4}(\g) \, \eta_{1,2}^\ast(\g) \, \biggl\{ \frac{1}{[3 - \eta_{1,2}^\ast(\g)]^2} - \frac{3}{8} \biggr\}, \\
\label{l2}
\lambda_2 & = & - 3H^\ast_{1,2,3,4}(\g).
\e
From these, the type of critical points
can be determined. \\
In the three sub-intervals for $\g$, for which critical points, different from the origin, exist, the situation is as follows. \\
Firstly, for $-\frac{8}{5} - \frac{32}{45} \sqrt{6} < \g < -1$ (i.e. $-3.3419 < \g < -1)$, there are five critical points (see Figure 2): the origin  $(\eta^\ast = 0, H^\ast = 0)$, the saddle point $A$ with coordinates $H^\ast_1(\g) > 0$ and $1.3670 < \eta_{1}^\ast(\g) < 3$ (at this point, $\lambda_1 > 0$ and $\lambda_2 < 0$); the saddle point $B$ with coordinates $H^\ast_3(\g) = - H^\ast_1(\g)< 0$ and $1.3670 < \eta_{1}^\ast(\g) < 3$ (at $B$, the eigenvalues change their signs:
$\lambda_1 < 0$ and $\lambda_2 > 0$); the stable node $C$ with coordinates  $H^\ast_2(\g) > 0$ and $\eta_{2}^\ast(\g) < 1.3670$ (at $C$ both eigenvalues are
negative); and the unstable node $D$ with coordinates  $H^\ast_4(\g) = - H^\ast_2(\g) < 0$ and $\eta_{2}^\ast(\g) < 1.3670$ (where both eigenvalues are
positive). \\
Next, for $-1 < \g < 0,$ there are only three critical points: the origin  $(\eta^\ast = 0, H^\ast = 0)$,
the saddle $E$ with coordinates $H^\ast_1(\g) > 0$ and $1.3670 < \eta_{1}^\ast(\g) < 3$ (where $\lambda_1 > 0$ and
$\lambda_2 < 0$) and the saddle $F$ with coordinates $H^\ast_3(\g) < 0$ and $1.3670 < \eta_{1}^\ast(\g) < 3$ (at which point the eigenvalues flip their signs:
$\lambda_1 < 0$ and $\lambda_2 > 0$). In this
regime, a cyclic Universe scenario appears again (but corresponding to a toy model with negative absolute temperature). \\
Finally, for  $0 < \g <  -\frac{8}{5} + \frac{32}{45} \sqrt{6}$ (i.e. $0 < \g < 0.1419)$, there are four equilibrium points [the origin  $(\eta^\ast = 0, H^\ast = 0)$ is no longer reachable as $\eta$ is always greater than 3]. These are: the stable node $G$ with coordinates $H^\ast_1(\g) > 0$ and $3 < \eta_{1}^\ast(\g) < 4.6330$ (both eigenvalues are negative there); the unstable node $K$ with coordinates $H^\ast_3(\g) < 0$ and $3 < \eta_{1}^\ast(\g) < 4.6330$ (where the eigenvalues are positive); the saddle $M$ with coordinates  $H^\ast_2(\g) > 0$ and $4.6330 < \eta_{2}^\ast(\g)$ (where $\lambda_1 > 0$ and $\lambda_2 < 0$); and  the saddle $N$ with coordinates  $H^\ast_4(\g) < 0$ and $4.6330 < \eta_{2}^\ast(\g)$ (where the eigenvalues are $\lambda_1 < 0$ and
$\lambda_2 > 0$). \\
At the origin $(\eta^\ast = 0, H^\ast = 0)$ of the phase portrait, $\lambda_1 = 0 = \lambda_2$ is a
double eigenvalue of the stability matrix $L(0, 0)$ --- as in the case of the virial real gas (note again that $\g$ must be negative for the origin to be reached).  To analyse the type of this critical point and its role, the help of the conserved quantity (first integral) $J$ will next be conjured.  \\
With the introduction of
\b
\label{tse}
c(\eta) = \eta \Bigl[ 1 + 3 \g \Bigl(\frac{1}{3-\g}-\frac{3 \eta}{8} \Bigr) \Bigr],
\e
the dynamical equations of the model (\ref{dyn2}) and (\ref{dyn1}) can be re-written as:
\b
\label{dyn2aa}
\dot{\eta} & = & - 3 H c(\eta), \\
\label{dyn1aa}
\dot{H} & = & - \frac{3}{2} H^2 + \frac{8 \xi}{3} [\eta -  c(\eta)].
\e
Next, a constant $J(H, \eta)$ is sought such that:
\b
0 = \frac{dJ}{dt} = \frac{\partial J}{\partial H} \dot{H} + \frac{\partial J}{\partial \eta} \dot{\eta}
\e
or, using the dynamical equations (\ref{dyn2aa}), (\ref{dyn1aa}),
\b
\frac{\partial J}{\partial H}
\Bigl[-\frac{3}{2} H^2 + \frac{8 \xi}{3} \eta - \frac{8 \xi}{3} c(\eta) \Bigr]
- 3 H c(\eta) \frac{\partial J}{\partial \eta}  = 0.
\e
One can change variables $\eta \rightarrow \tilde{\eta}(\eta)$ via:
\b
\label{changevari}
- c(\eta) \frac{\partial}{\partial \eta} = \tilde{\eta} \frac{\partial}{\partial \tilde{\eta}}.
\e
Then $J[H, \eta(\tilde{\eta})]$ satisfies:
\b
\frac{\partial J}{\partial H}
\Bigl[-\frac{3}{2} H^2 +
\frac{8 \xi}{3} \eta(\tilde{\eta}) - \frac{8 \xi}{3} c[\eta(\tilde{\eta})]\Bigr]
+ \frac{\partial J}{\partial \tilde{\eta}} [3 H \tilde{\eta}] = 0.
\e
Thus $J[H, \eta(\tilde{\eta})]$ can be found as a solution of the system:
\b
\label{ggg1}
\frac{\partial J}{\partial \tilde{\eta}} & = & -\frac{3}{2} H^2 +
\frac{8 \xi}{3} \eta(\tilde{\eta}) - \frac{8 \xi}{3} c[\eta(\tilde{\eta})], \\
\label{ggg2}
\frac{\partial J}{\partial H}
& = & - 3 H \tilde{\eta}.
\e
Integrating (\ref{ggg1}) with respect to $\te$ gives:
\b
J = -\frac{3}{2} H^2 \te +
\frac{8 \xi}{3}
\int \Bigl[ \eta(\tilde{\eta}) - c[\eta(\tilde{\eta})] \Bigr] d\te.
\e
Using
\b
\label{razd}
d\te = - \frac{\te}{c(\eta)}d\eta,
\e
the above becomes:
\b
\label{jay}
J = -\frac{3}{2} H^2 \te +
\frac{8 \xi}{3}
\int d(\eta \te)
 = \Bigl( -\frac{3}{2} H^2 +
\frac{8 \xi}{3} \eta \Bigr)\te.
\e
To find $ \te(\eta)$, equation (\ref{razd}) can be integrated (it is in separate variables):
\b
\ln \te = - \int \frac{d \eta}{c(\eta)} + \mbox{const}.
\e
The integration constant is irrelevant and can be ignored  [it is obvious from the definition (\ref{changevari}) that $\te$ is defined modulo a multiplicative constant].  Thus:
\b
\te =
e^{\int \frac{(\eta-3)d\eta}{\eta (\hat{a} \eta^2 + \hat{b} \eta + \hat{c})}} =
\vert \eta \vert^{\frac{-3}{\hat{c}}} \,\,
\vert \eta - \eta_1^* \vert
^{\frac{3}{2\hat{c}} - \frac{3\hat{b}+2\hat{c}}{2\hat{c}\sqrt{\hat{b}^2-4\hat{a}\hat{c}}}} \,\,
\vert \eta - \eta_2^* \vert
^{\frac{3}{2\hat{c}} + \frac{3\hat{b}+2\hat{c}}{2\hat{c}\sqrt{\hat{b}^2-4\hat{a}\hat{c}}}},
\e
where: $\hat{a} = 9 \g/8$, $\hat{b}= -(27 \g / 8 + 1),$  $\hat{c}=3(1+\g)$, and $\g$ is such that $\hat{b}^2 - 4 \hat{a}\hat{c} > 0$ always.   \\
One can immediately notice from the above change of variables that the Poisson bracket of the model is singular for all values of $\gamma$ in the interval $(-3.3419,-1)$ at the equilibrium points  $0$ and the sink and the source at $\eta_{2}^*$ (the dependence on the temperature $\gamma$ is via $\hat{a}$, $\hat{b}$, and $\hat{c}$). These three critical points are mapped to infinity in the canonical coordinates.  At the two saddles at $\eta_{1}^*$, the Poisson bracket is singular for $-3.3419 < \gamma \le -3.2000$ and regular for $-3.2000 < \gamma < -1$.\\
Secondly, this fact explains the appearance of the stable node $C$ with coordinates  $H^\ast_2(\g) > 0$ and $0 < \eta_{2}^\ast(\g) < 1.3670$
when $-\frac{8}{5} - \frac{32}{45} \sqrt{6} < \g < -1$ (i.e. $-3.3419 < \g < -1)$ and the appearance of the stable node $G$ with coordinates $H^\ast_1(\g) > 0$ and $3 < \eta_{1}^\ast(\g) < 4.6330$ when $0 < \g <  -\frac{8}{5} + \frac{32}{45} \sqrt{6}$ (i.e. $0 < \g < 0.1419)$. As mentioned earlier, planar Hamiltonian systems can only have centres and saddles. However, as the Hamiltonian structure is lost on the equilibrium points, the emergence of a stable node is not a violation of the arguments presented earlier. In Hamiltonian variables, some of the equilibrium points are moved to infinity, however it is still possible to have nodes as equilibrium points in terms of the original physical variables.  \\
From (\ref{jay}), it can be seen that the parabola
\b
\label{par}
\eta = \frac{9}{16 \xi} H^2
\e
is a special curve: $J$ vanishes on it. \\
Also, all five equilibrium points (when $-3.3419 < \g < -1$), all three equilibrium points (in the case when $-1 < \g < 0$), and all four equilibrium points (when $0 < \g < 0.1419$) are on this parabola.
\begin{center}
\includegraphics[width=6cm]{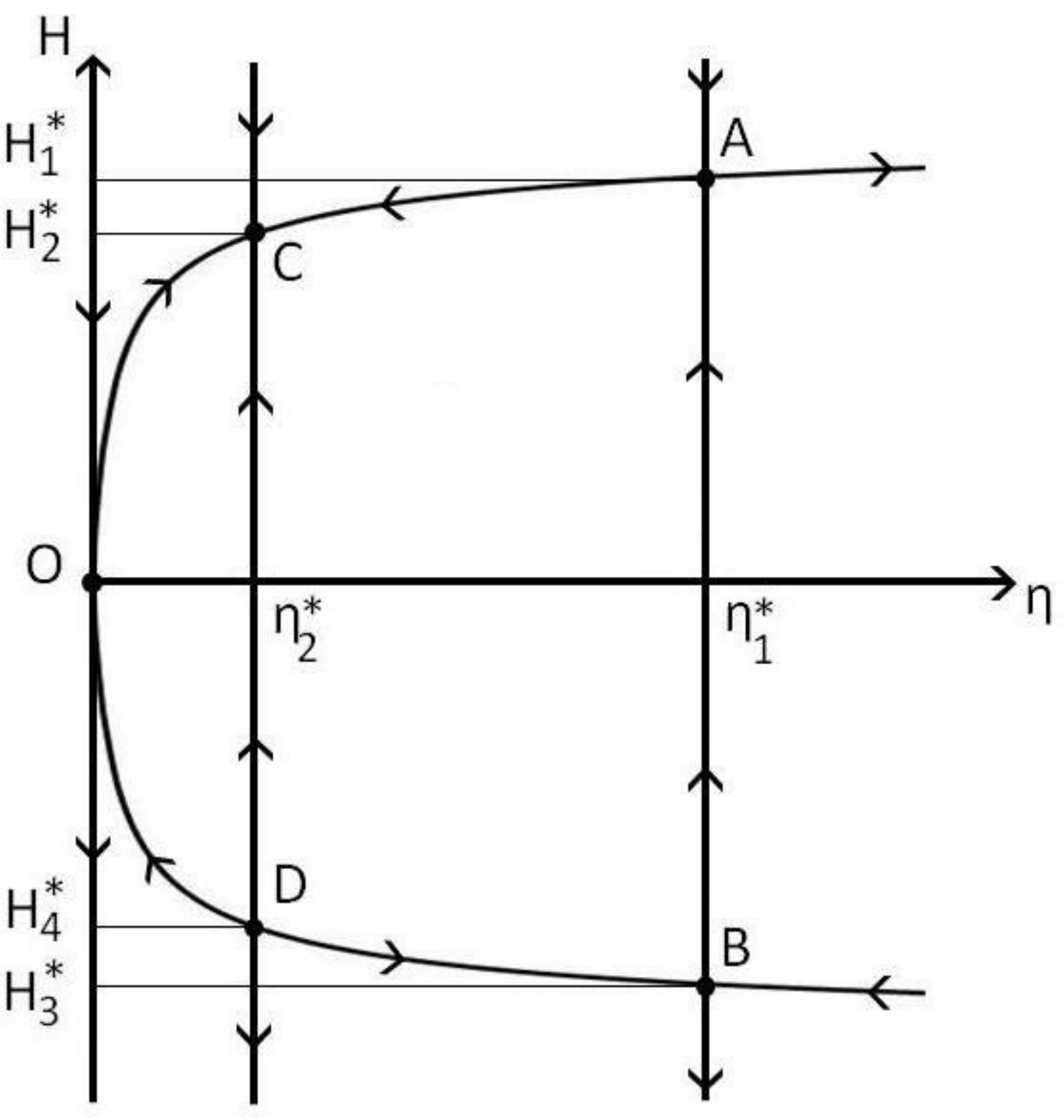}
\noindent
\parbox{100mm}{\footnotesize {\bf Figure 2}: Critical points for the van der Waals gas for  $-3.3419 < \g < -1$. The parabola $\eta=(9H^2)/(16 \xi)$ contains all five critical points.  }
\vskip.4cm
\end{center}
The dynamical equations (\ref{dyn2}) and (\ref{dyn1}) are, obviously, satisfied on the parabola: differentiating (\ref{par}) with respect to time, yields $\dot{\eta} = 9/(8 \xi) H \dot{H}$ and substituting here (\ref{dyn1}), results in (\ref{dyn2}). Separately, expressing $H$ as $\pm (4/3) \sqrt{\xi \eta}$ along the parabola and substituting this into $\dot{H} = (8 \xi)/(9 H) \dot{\eta}$, gives $\dot{H} = \pm (2 \xi \dot{\eta})/(3 \sqrt{\xi \eta})$.  Substituting the dynamical equation (\ref{dyn1}) into the latter yields
\b
\dot{\eta} = \mp 4 \eta \sqrt{\xi \eta} \Bigl[ 1 + 3 \g \Bigl( \frac{1}{3-\eta} - \frac{3}{8} \eta \Bigr) \Bigr]
\e
along the parabola. \\
Substituting $H^2 = (16 \xi/9) \eta$ into the right-hand side of the dynamical equation (\ref{dyn2}) gives:
\b
\dot{H} = - \frac{\frac{8}{3} \xi \eta}{3 - \eta}
\Bigl[ 3(\g+1) - \Bigl( \frac{27}{8} \g + 1 \Bigr) \eta + \frac{9 \g}{8} \eta^2 \Bigr].
\e
The zeroes of the quadratic expression in the square brackets are exactly $\eta_{1,2}^*$, given by (\ref{eta_roots}). Thus:
\b
\dot{H} = - \frac{3 \xi \g \eta}{3-\eta} (\eta - \eta_1^*)(\eta - \eta_2^*).
\e
Returning to $H$ on the right-hand side, results in:
\b
\dot{H} = - \Bigl( \frac{27}{64 \xi}\Bigr)^2 \frac{\g}{1-\frac{3 H^2}{16 \xi}} H^2 [H^2 - (H_1^*)^2] [H^2 - (H_2^*)^2]
\e
or
\b
 \frac{1-\frac{3 H^2}{16 \xi}}{H^2 [H^2 - (H_1^*)^2] [H^2 - (H_2^*)^2]} dH =
- \Bigl( \frac{27}{64 \xi}\Bigr)^2 \g dt.
\e
Expressing the left-hand side in partial fractions, gives:
\b
\label{tukaa}
\biggl[ \frac{m}{H^2} + \frac{n_1}{H^2 - (H_1^*)^2} + \frac{n_2}{H^2 - (H_2^*)^2} \biggr] dH =
- \Bigl( \frac{27}{64 \xi}\Bigr)^2 \g dt,
\e
where:
\b
m & = & \frac{1}{(H_1^* H_2^*)^2}, \\
n_1 & = & \frac{1-\frac{3 (H_1^*)^2}{16 \xi}}{(H_1^*)^2 [(H_1^*)^2 - (H_2^*)^2]}, \\
n_2 & = &  \frac{1-\frac{3 (H_2^*)^2}{16 \xi}}{(H_2^*)^2 [(H_2^*)^2 - (H_1^*)^2]}.
\e
Using:
\b
\frac{1}{H^2 - (H_{1,2}^*)^2}
= \frac{1}{2 H_{1,2}^*}\Bigl( \frac{1}{H - H_{1,2}^*} -
\frac{1}{H - H_{3,4}^*} \Bigr)
\e
and integrating (\ref{tukaa}), results in:
\b
-\frac{m}{H} + \frac{n_1}{2 H_1^*} \ln \Biggl|\frac{H - H_1^*}{H - H_3^*} \Biggr|+ \frac{n_1}{2 H_2^*} \ln \Biggl| \frac{H - H_2^*}{H - H_4^*} \Biggr|= - \Bigl( \frac{27}{64 \xi}\Bigr)^2 \g (t - t_O).
\e
(The terms involving $H_0 = H(t_0)$ from the integrals on the left-hand side have been absorbed into $t_O$ on the right-hand side.) \\
Thus, along the parabola (\ref{par}), the origin and equilibrium points $(\eta_{1,2}^*, H_{1,2}^*)$ are reachable in infinite time ($t \to \infty$), while
 $(\eta_{1,2}^*, H_{3,4}^*)$ are reachable in $t \to -\infty$. \\
When $-1 < \g < 0$, there are only three equilibrium points: the origin, $(\eta_{1}^*, H_{1}^*)$ (both reachable in $t \to \infty$) and $(\eta_{1}^*, H_3^*)$ (reachable in $t \to - \infty$). \\
Along the vertical line $\eta = 0$, the dynamical equations reduce to $\dot{H} = - (3/2) H^2$. Integration gives $1/H = (3/2)(t-t_0)$. Thus, reaching the origin, i.e. $H \to \pm 0$, takes time $t \to \pm \infty$. \\
Consider next moving along either of the vertical lines $\eta = \eta_{1,2}^* =$ const towards any of the remaining four (or two, when $-1 < \g < 0$) equilibrium points  $(\eta_{1,2}^*, H_{1,2}^*)$ and $(\eta_{1,2}^*, H_{3,4}^*)$ [all of which lie on the parabola (\ref{par}) where
$\eta_1^* = 9/(16 \xi) (H_{1,3}^*)^2$ and $\eta_2^* = 9/(16 \xi) (H_{2,4}^*)^2$]. Along $\eta = \eta_{1,2}^*$, the dynamical equations reduce to
\b
\dot{H} = - \frac{3}{2} H^2 - 8 \xi \g \eta_{1,2}^* \Bigl(\frac{1}{3 - \eta_{1,2}^*} - \frac{3}{8} \eta_{1,2}^* \Bigr).
\e
The term in the brackets, call it $\mu$, is equal to
$(1/8) [3(\eta_{1,2}^*)^2 - 9 \eta_{1,2}^* +8]/(3-\eta_{1,2}^*)$. Using (\ref{hash_roots}), it follows that $(H_{1,2}^*)^2 = (H_{3,4}^*)^2 = -(16/9) \xi \eta_{1,2}^* \mu$. Substituting here $\eta_1^* = 9/(16 \xi) (H_{1,3}^*)^2$ and $\eta_2^* = 9/(16 \xi) (H_{2,4}^*)^2$, it immediately gives $\mu =-1/(3\g)$. \\
Therefore:
\b
\frac{dH}{H^2 - (H_{1,3}^*)^2} = -\frac{3}{2} dt
\e
at the two equilibrium points along $\eta = \eta_{1}^*$ (and similarly along $\eta = \eta_{2}^*$).
Integrating gives:
\b
\frac{1}{2 H_{1,3}^*} \, \ln \Biggl|\frac{H - H_{1,3}^*}{H-H_{3,1}^*}\Biggr| = - \frac{3}{2} (t - t_0).
\e
Therefore, if $H$ tends to $H_1^*$ from above, the equilibrium point
$(\eta_1^*, H_1^*)$ is reachable in time $t \to -\infty$. This point is not reachable at all if $H$ tends to $H_1^*$ from below (time becomes purely imaginary). \\
If  $H$ tends to $H_3^*$ from above, the equilibrium point
$(\eta_1^*, H_3^*)$ is not reachable (imaginary time) and if $H$ tends to $H_3^*$ from below, then this equilibrium point is reachable in time $t \to -\infty$.  \\
Similar analysis holds for moving along $\eta = \eta_2^*$. \\
Finally, the trajectories within the parabola will be addressed (see Figure 3). To do so, the dynamical equations (\ref{dyn2}) and (\ref{dyn1}) will be linearised near point $(H=0, \eta = \eta_0)$.
Using (\ref{jay}), one can express $-(3/2) H^2$ as $J/\te - (8/3) \xi \eta$ and substituting this into (\ref{dyn1aa}) gives:
\b
\dot{H} = \frac{J}{\te(\eta)} - \frac{8}{3} \xi c(\eta).
\e
Expand this at $\eta_0 + r$ and retain up to and including the term linear in the small $r$:
\b
\dot{H}(\eta_0 + r) & = & J \Bigl[ \frac{1}{\te(\eta_0)} + \Bigl( \frac{d}{d\eta} \frac{1}{\te(\eta)} \Bigr)_{\eta = \eta_0} r \Bigr]
- \frac{8}{3} \xi [c(\eta_0) - c'(\eta_0) r] \nonumber \\
& = & \Bigl[ \frac{J}{\te(\eta_0)} - \frac{8}{3} \xi c(\eta_0) \Bigr]
+ \Bigl[ J a_1 -  \frac{8}{3} \xi c'(\eta_0)\Bigr]r,
\e
where $a_1 = (d/d\eta)[1/\te(\eta)]_{\eta = \eta_0}$. \\
In the first square brackets of the last equality, $\eta_0$ could be chosen in such way that this term becomes zero. Namely:
\b
\label{e0}
\dot{H}(\eta_0) = \Bigl[ \frac{J}{\te(\eta_0)} - \frac{8}{3} \xi c(\eta_0) \Bigr] = 0.
\e
It is possible to do so and (\ref{e0}) will be the equation defining $\eta_0$. Obviously, $\eta_0$ depends on the value of $J$, that is, on the initial conditions. As it will turn out, $\eta_0$ will be the centre of the elliptical trajectories confined within the parabola (\ref{par}).
The linearised dynamical equation for the Hubble's parameter near point $(0, \eta_0)$ is therefore:
\b
\dot{H}(\eta_0 +r) = \Bigl[ J a_1 - \frac{8}{3} \xi c'(\eta_0) \Bigr] r = - l_1 r,
\e
where $l_1= - [J a_1 - (8/3) \xi c'(\eta_0)] = $const.
\begin{center}
\includegraphics[width=6cm]{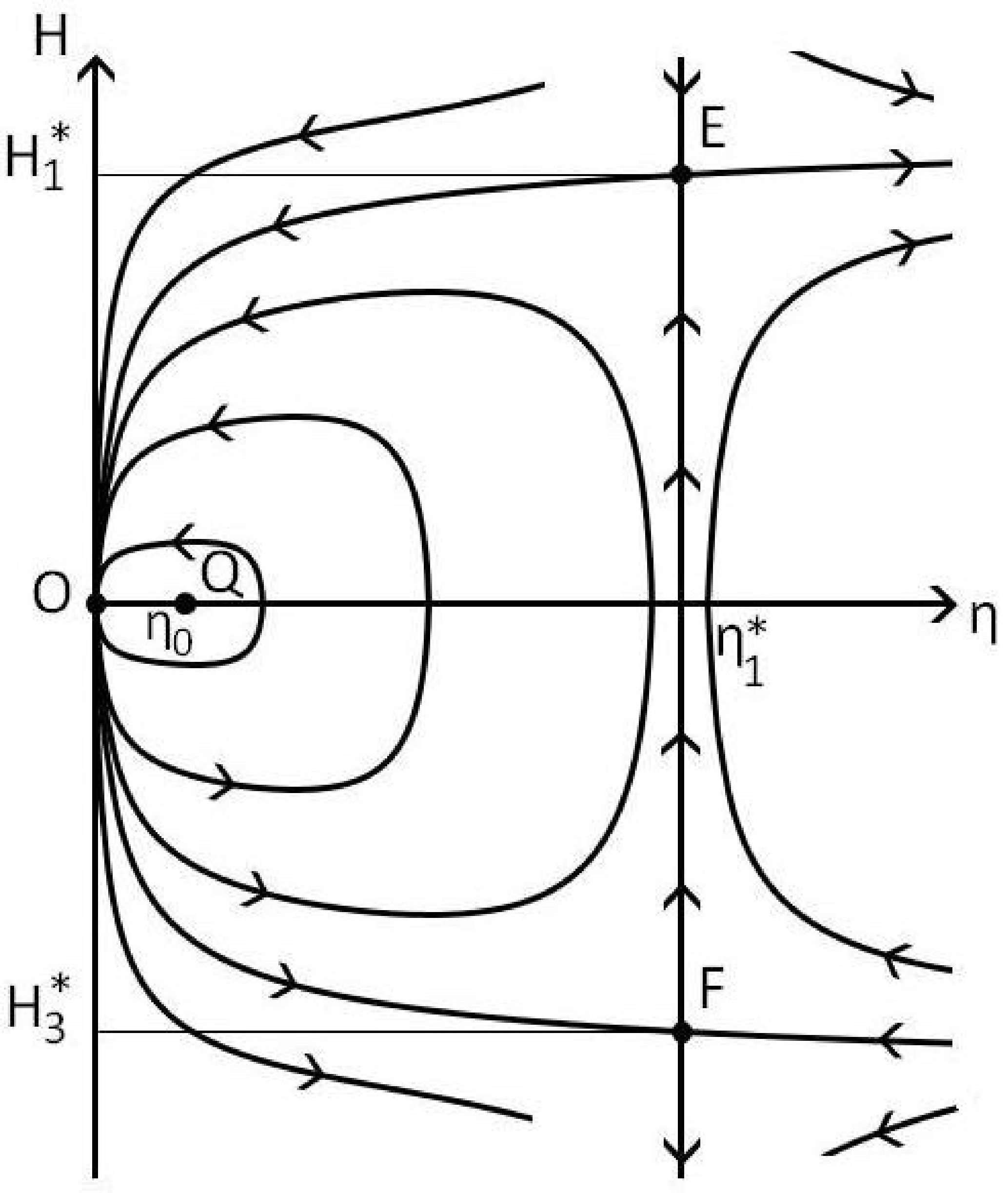}
\noindent
\parbox{100mm}{\footnotesize {\bf Figure 3}: Closed trajectories for van der Waals gas for $-1 < \g < 0$.}
\vskip.4cm
\end{center}
The linearised second dynamical equation is
\b
\dot{\eta}(0+h, \eta_0 + r) =
- 3 c(\eta_0) h = l_2 h,
\e
where $l_2 = - 3 c(\eta_0)$ is a constant and $h$ is infinitesimal. \\
Using the defining equation for $\eta_0$, (\ref{e0}), one finds:
\b
l_1 = \frac{8}{3} \xi [1 - c'(\eta_0)].
\e
The constants $l_1$ and $l_2$ must have the same signs for closed elliptical trajectories to exist in the phase-plane  (as it will be shown further). \\
Using the definition of $c(\eta)$, (\ref{tse}), it follows that:
\b
l_1 = - 6 \xi \g \biggl( \frac{1- \eta_0}{3-\eta_0}\biggr)^2 (\eta_0-4).
\e
When $\g$ is negative, as already discussed, all equilibrium points (five, when $-8/5 - $\linebreak$(32/45) \sqrt{6} < \g < -1$ and three when $-1 < \g < 0)$) are, together with $\eta_0$, to the left of 3. Thus $l_1$ is always negative.  \\
However, if $\g$ is in the region $0 < \g < -\frac{8}{5} + \frac{32}{45} \sqrt{6}$ (in which case all four critical points are to the right of 3), then $l_1$ is negative only if $\eta_0 > 4$ and positive if $3 < \eta_0 < 4$. \\
On the other hand,
\b
l_2 = - 3 \Bigl[ \frac{9}{8}\g\eta_0^2 - \Bigl(\frac{27}{8}\g+1\Bigr) \eta_0 + 3 (\g+1)\Bigr].
\e
The zeroes of the expression in the square brackets are exactly the roots $\eta_{1,2}^*$, as can be seen from (\ref{roots}). \\
If $\g$ is negative [$-8/5 - (32/45) \sqrt{6} < \g < 0$], then $l_2$ is negative for all $\eta_0$ between the critical
points $\eta_{1,2}^*$. Otherwise, when $0 < \g < -8/5 + (32/45) \sqrt{6}$, then $l_2$ is negative for $3 < \eta_0 < \eta_2^*$ and for $\eta_0 > \eta_1^*$. \\
The case of positive $\g$ will not be analysed
further --- simply because of the presence of the stable node $G$ with coordinates $H^\ast_1(\g) > 0$ and $3 < \eta_{1}^\ast(\g) < 4.6330$
which "sucks in" all trajectories and does not allow the existence of closed curves on the phase portrait. \\
Next, elliptical trajectories are sought in the phase-plane for the case of negative $\g$:
\b
r & = & A \sin \omega t, \\
h & = & B \cos \omega t.
\e
Thus, $\dot{r} = (A \omega/B)h$ and $\dot{h} = - (B \omega / A)r$. Comparing this to the linearised dynamical equations, $\dot{r} = l_2 h$ and $\dot{h} = -l_1 r$, yields $l_1 = B \omega/ A$ and $l_2 = A \omega / B$. Thus $l_1 l_2 = \omega^2 = -8\xi c(\eta_0)[1-c'(\eta_0)]$. \\
The equation of the ellipse is:
\b
\Bigl(\frac{r}{A}\Bigr)^2 + \Bigl(\frac{h}{B}\Bigr)^2 = 1.
\e
Clearly, $A = \eta_0$ and $B = \omega \eta_0/l_2$. Thus:
\b
r & = & \eta_0 \sin \omega t, \\
h & = & \frac{\omega \eta_0}{l_2} \cos \omega t.
\e
One can also relate $J$ to $l_1$ and $l_2$. To do so,
consider (\ref{jay}) in the form
\b
-\frac{3}{2} H^2 = \frac{J}{\te} - \frac{8}{3}\xi \eta
\e
and expand near $(\eta_0, 0)$. One gets:
\b
\frac{3}{2} H^2 = - J (a_0 + a_1 r + a_2 r^2) - \frac{8}{3}\xi (\eta_0 + r).
\e
Here, $a_0 = 1/\te(\eta_0)$,
$a_1 = (d/d\eta) [1/\te(\eta)]_{\eta=\eta_0}$, and $a_2 = (1/2) (d^2/d\eta^2) [1/\te(\eta)]_{\eta=\eta_0}$. \\
Given that $a_1 = 1/ [\te(\eta_0) c(\eta_0)]$, the term $(8/3) \xi - J a_1$ vanishes due to the defining equation for $\eta_0$, (\ref{e0}). This leaves:
\b
\frac{3}{2} h^3 + J a_2 h^2 = \mbox{const. }
\e
Upon comparing this to $l_1 r^2 + l_2 h^2 =$ const, it allows to express $l_1$ in terms of $l_2$ through the boundary conditions (via $J$):
\b
l_1 = \frac{2 a_2 J}{3} l_2.
\e
One can make the following observation: $\eta=\eta_0, \,\, H=0$ is a trivial solution of the linearised system, but it is not an equilibrium point for the original nonlinear system. Moreover, $\eta_0$ depends on the initial conditions through the value of the first integral $J$. The trajectories do not pass through this point --- for the linearised system these trajectories are ellipses with centres at that point.
\\
The trajectories (ellipses in the linearised case) however always contain the origin  $(\eta^\ast = 0, H^\ast = 0)$ as an equilibrium point.    From the point of view of the dynamical systems theory (Poincar\'e--Bendixson Theorem), the situation is that of trajectories, which are trapped within the region between the origin $(\eta^\ast = 0, H^\ast = 0)$, the parabola $\eta = (9H^2)/(16\xi)$ and the vertical line $FE$ ($\eta=\eta_1^*$). There is no stable equilibrium in this region and the trajectories are in the form of homoclinic orbits through the origin $(\eta^\ast = 0, H^\ast = 0)$ and these are ellipses in the linear approximation. The boundary of the trapping region itself is a heteroclinic orbit passing through the equilibria $(\eta^\ast = 0, H^\ast = 0)$ and the two saddles: $E$ with coordinates $(\eta_1^*, \, H_1^*)$ and $F$ with coordinates ($\eta_1^*, H_3^*)$. \\
The second integrals for the van der Waals model are $M_1 = \eta - (9H^2)/(16\xi)$ (conserved on the parabola), $M_2 = \eta$ (existing when $-8/5 - (32/45) < \g < 0$ and conserved along $\eta^* = 0$), $M_3 = \eta - \eta_1^*$ (conserved along $\eta = \eta_1^*$) and, finally, when $-8/5 - (32/45) \sqrt{6} < \g < -1$ or  $0 < \g < -8/5 + (32/45) \sqrt{6}$  the second integral $M_4 = \eta - \eta_2^*$ is conserved on $\eta = \eta_2^*$.

\section{Discussion}
A large class of cosmological models can be formulated as a dynamical system of two autonomous ordinary differential equations. The nonlinear dynamics in two dimensions is both relatively simple and very well studied. It is particularly simple when a global first integral exists. \\
Apparently, certain classes of cosmological models admit a global conserved quantity ---  illustrated with the presented examples. In addition, this conserved quantity can serve as a Hamiltonian for a canonical Hamiltonian formulation of the evolution equations.
Moreover, in the case of more complicated Hamiltonian systems, one way of doing consistent approximations is to approximate the Hamiltonian, which is a scalar function, rather than working with each equation.
The canonical Hamiltonian formulation necessitates canonical coordinates, which could be obtained from the physical ones with a nonlinear change (transformation) of variables. It turns out that the coordinate transformation to canonical variables can be singular at some of the possible equilibrium points. This phenomenon is most certainly related to the fact that Hamiltonian systems in two dimensions allow only for centre and saddle type equilibria. Thus, any other (node-type) of equilibrium point is mapped to infinity under the coordinate change from physical to canonical variables. Thus, the advantage of the canonical coordinates from practical point of view is debatable, but one should bear in mind that a stable sink --- any static equilibrium --- is not reachable in finite time. This indicates that the sinks are not a topological feature of the model(s), but rather, an artefact reflecting the particular choice of physical coordinates.  In contrast, the closed orbits and the saddles determine the essential behaviour of the system, since these always appear in the spectrum of the (linearised) Hamiltonian formulation. \\
In addition to the global first integral, there are often special (second) integrals, defined and conserved on a lower-dimensional manifold (lines or curves) in the two-dimensional phase space. The conserved quantities lead to the existence of stable periodic solutions (closed orbits) which are models of a cyclic Universe. The integrals also allow for explicit solutions (as functions of time $t$) on some of the system trajectories and thus for a deeper understanding of the underlying physics. The periodic solutions are special ones and their stability is established rigorously with the help of the first integral (which in the right variables produces the Hamiltonian). Moreover, the parameters of the closed orbits are related to the value of the first integral (and thus to the initial conditions) as well as the parameters of the system. The nature of the closed orbits is also established - a centre in the first model and a homoclinic orbit in the second one. In the limit, heteroclinic orbits are possible.

\vskip.5cm
\noindent
Dedicated to the memory of Nadejda Vassileva Manova--Prodanova (14.01.1926 -- \linebreak 11.04.2016).

\end{document}